\documentclass[final]{aipproc}
\layoutstyle{8x11single}

\usepackage{graphicx}
\usepackage{lmodern}
\usepackage{url}

\begin{document}

\title{Top Quark Production Measurements at ATLAS}


\author{R. Calkins on behalf of the ATLAS collaboration\footnote{Copyright CERN for the benefit of the ATLAS collaboration}}{
  address={Northern Illinois University, Northern Illinois Center for Accelerator and Detector Development (NICADD), DeKalb, IL 60115, United States}
}

\begin{abstract}
An overview of top quark production measurements using the ATLAS detector at the LHC is presented. Using 35 pb$^{-1}$ of data, we measured the $t\overline{t}$ cross-section in the lepton+jets channel to 13\% precision and set limits on the cross-section for the all hadronic decay channel\cite{ljetbtag,allhad}. The measurement in the dilepton channel was performed using 0.70 fb$^{-1}$ of data and was combined with the lepton+jets channel result for our most precise $t\overline{t}$ cross-section measurement of $\sigma_{pp \rightarrow t\overline{t}+X}=176\pm 5(\mathrm{stat})^{+13}_{-10}(\mathrm{syst})\pm 7(\mathrm{lumi})$\cite{dilep,combo}. Single top production was also measured in the $t$-channel using  0.70 fb$^{-1}$ of data\cite{tchan}.

\end{abstract}

\maketitle

%%%%%%%%%%%%%%%%%%%%%%%%%%%%%%%%%%%%%%%%%%%%
%% MAINMATTER
%%%%%%%%%%%%%%%%%%%%%%%%%%%%%%%%%%%%%%%%%%%%

\section{Top Quark Production and Decay}

Due to its high center of mass energy, the LHC is expected to produce large amounts of top quarks. The measurement of the production rate is an important milestone in the LHC's physics program.  At the LHC, the majority of top quarks are produced in pairs from gluon-gluon fusion but if there is a yet undiscovered heavy particle, it may decay into top pairs and enhance the cross-section.  In addition to pair production, top quarks can be produced individually through electroweak processes. Single top production was previously observed at CDF and D0\cite{cdf_s,d0_s}. 
Nearly 100\% of the time a top quark will decay into a $b$-quark and a $W$-boson. Experimentally, $t\overline{t}$ events are classified based on how the $W$-bosons decay. If both $W$-bosons decay leptonically, it is called the dilepton channel and if just one of the $W$-bosons decays leptonically, it is referred to as the lepton+jets channel.  For the case when both decay hadronically, it is referred to as the all hadronic channel. For the rest of this article, the term lepton only refers to electrons and muons.

\section{Lepton+jets Channel}

Often called the Golden Channel, the lepton+jets channel has a sizeable branching ratio plus the advantage of having a final state lepton to trigger the event. For this analysis, events are triggered by unprescaled single lepton triggers and a cut on $p_T$ is applied above the trigger threshold so that the efficiency is nearly constant as a function of $p_T$.
 We require a final state topology of one lepton (matched to the trigger object) and at least three jets.
Additionally for the electron channel, ${{E}_T}^{miss}>35$ GeV and transverse $W$-boson mass($m_T$) $>25$ GeV are added cuts and for the muon channel, ${{E}_T}^{miss}>20$ GeV and ${{E}_T}^{miss}+m_T>60$ GeV are required conditions to pass the event selection.

The lepton+jets channel suffers from two large backgrounds, QCD jets faking leptons and $W$+jets. Monte Carlo is unreliable in predicting the QCD background
so data-driven methods are used instead.  The contribution from W+jets is estimated from LO Monte Carlo samples.  Multivariate methods were used to extract the $t\overline{t}$ cross-sections from the data samples.   
Three uncorrelated variables, pseudo-rapidity of the lepton ($\eta_{lepton}$), lepton charge ($q_{lepton}$) and the exponential of -8$\times$ aplanarity ($exp(-8\times Aplanarity)$), were selected and individual likelihoods were derived for each.
 These were multiplied together to form a total likelihood and this distribution was fitted to extract the cross-section. This method measured a $t\overline{t}$ cross-section of
$\sigma_{pp \rightarrow t\overline{t}+X}=171\pm 17(\mathrm{stat})^{+20}_{-17}(\mathrm{syst})\pm 6(\mathrm{lumi})$ pb \cite{ljetno}
. A separate measurement was also performed utilizing $b$-tagging information\cite{ljetbtag}. A discrimination variable, the average $b$-tagging weight of the two highest weighed jets in the event, was introduced and the charge variable was replaced by a $H_T$ based variable. This analysis is slightly more precise and returned a result of 
$\sigma_{pp \rightarrow t\overline{t}+X}=186\pm 10(\mathrm{stat})^{+21}_{-20}(\mathrm{syst})\pm 6(\mathrm{lumi})$ pb,
 which is consistent with the other analysis.  It is important to note that the two analyses are sensitive to different systematics.  The analysis using $b$-tagging is sensitive to $Wbb$, $Wcc$ and $Wc$ theory uncertainties while being less sensitive to QCD modeling than the other analysis.

\section{Dilepton Channel}

Having the smallest branching ratio, the dilepton channel makes up for it by having a generally higher purity. Requiring two oppositely charged leptons in
 the final state greatly reduces the QCD background.  For same flavor channels, $ee$ and $\mu\mu$, Drell-Yan processes contribute a substantial number of events.  To
reduce this number, events where $m_{ll}$ falls within a 10 GeV window around the Z-peak are rejected. To avoid light resonances, a cut of $m_{ll}$ >15 GeV is 
applied. For the $e\mu$ channel, the cut on ${{E}_T}^{miss}$ is dropped and a cut on $H_T$ > 130 GeV is applied instead.

The number of jets for all channels and after all selection cuts are applied is shown in Fig. 1. At least two jets are required to improve the $S/B$ ratio. 
As shown in Fig. 1, requiring at least 1 $b$-tagged jet can further improve this at the cost of statistics.  A `cut and count` method was used and the cross-section is extracted using a profile likelihood ratio.  For the default selection,  we measured 
$\sigma_{pp \rightarrow t\overline{t}+X}=171\pm 6(\mathrm{stat})^{+16}_{-14}(\mathrm{syst})\pm 8(\mathrm{lumi})$ pb
and when requiring a $b$-tagged jet, 
$\sigma_{pp \rightarrow t\overline{t}+X}=177\pm {6(\mathrm{stat})^{+17}_{-14}}(\mathrm{syst})^{+8}_{-7}(\mathrm{lumi})$ pb
.

\begin{figure}[dilepplot]
  \begin{minipage}[b]{0.3\linewidth}
    \centering
    \includegraphics[height=.2\textheight]{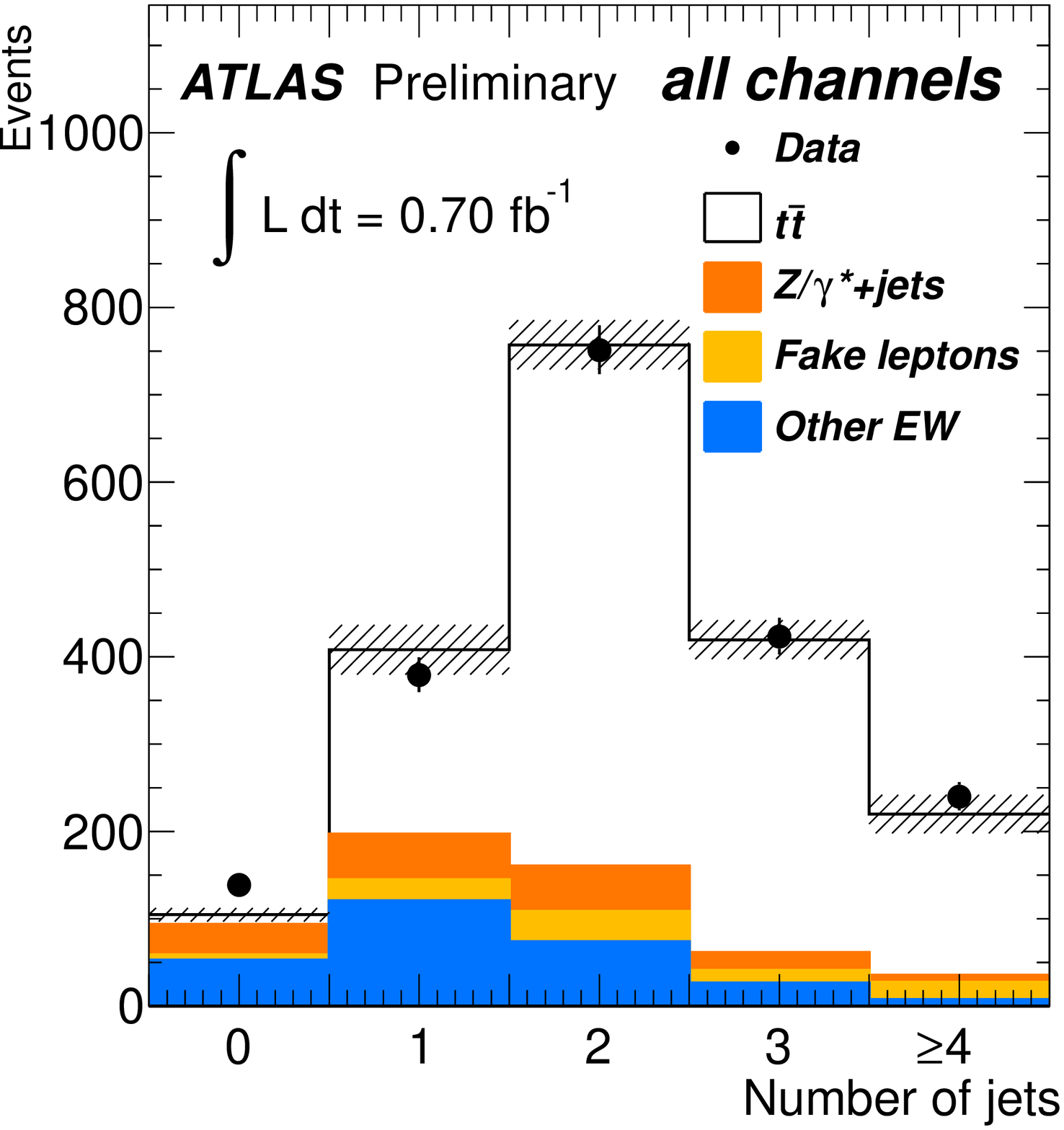}
    \caption{The caption for figure 1}
    \label{dilepfig1}
  \end{minipage}
  \hspace{0.0cm}
  \begin{minipage}[b]{0.3\linewidth}
    \centering
    \includegraphics[height=.2\textheight]{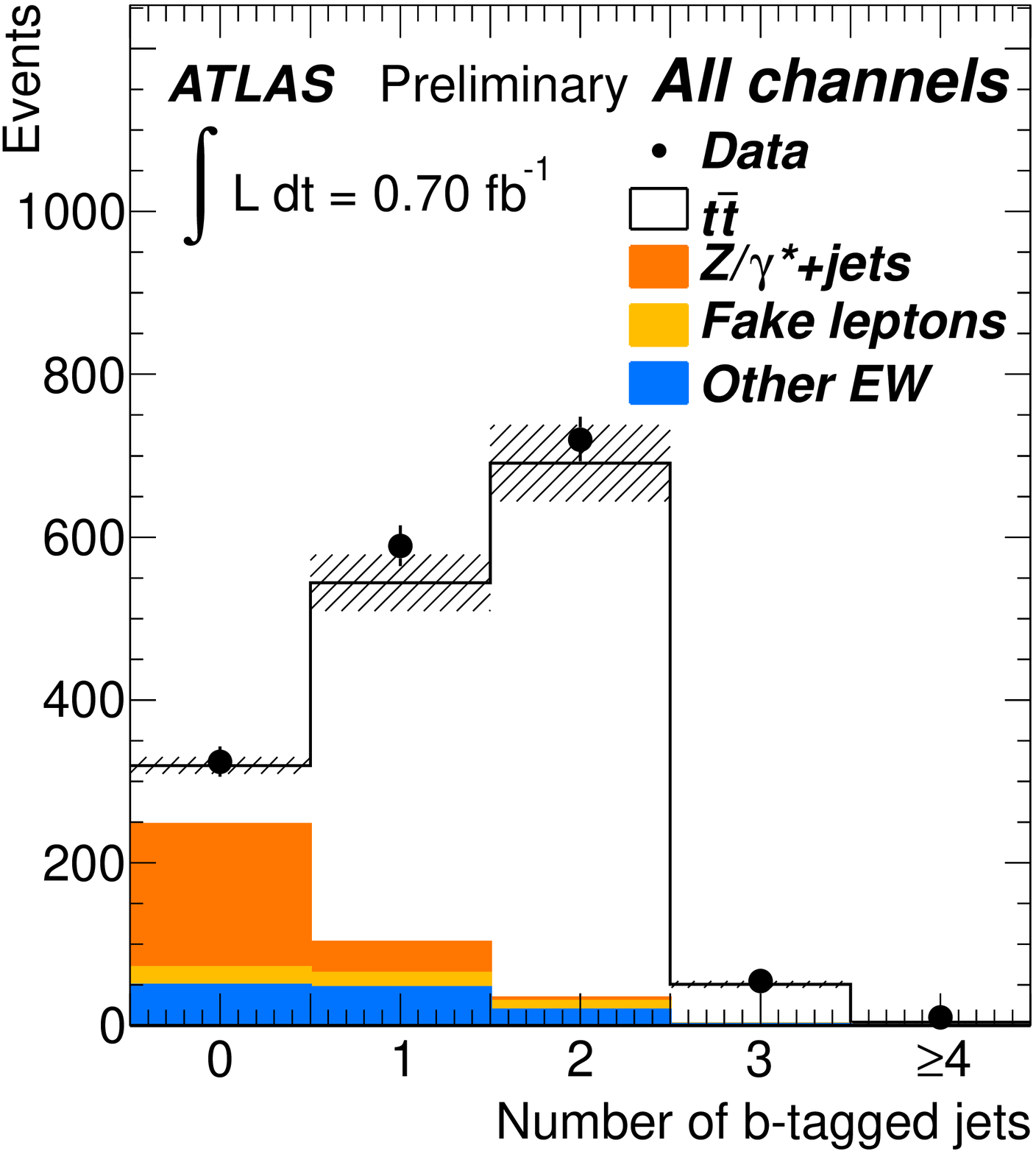}
    \caption{The number of jets in events passing the dilepton event selection (left) and the number of $b$-tagged jets (right). }
    \label{dilepfig2}
  \end{minipage}
\end{figure}

\section{All Hadronic Channel}

The all hadronic channel has the largest branching ratio but also the largest background since the $t\overline{t}$ signal is overwhelmed by the enormous QCD
cross-section. Since there is no lepton, a trigger requiring 4 high $p_T$ jets is used instead and 6 jets reconstructed offline with $p_T >60$ GeV are required. 
Since two of the jets are expected to originate from $b$-quarks, two $b$-tagged jets are required. Since the mass of the top and $W$-boson is known, this
knowledge is used to derive a discriminating variable. The variable, $\chi^{2}$, is defined as :
\begin{equation}
\chi^{2} = \displaystyle\sum\limits_{i=1}^{2} (\frac{m_{jjb}^{i}-m_{top}}{\sigma_{top}})^{2} +  (\frac{m_{jj}^{i}-m_{W}}{\sigma_{W}})^{2} 
\end{equation}
All combinations in an event are examined and the smallest $\chi^{2}$ value is taken (see Fig. 2). The distribution is then fitted to determine the $t\overline{t}$
cross-section (Fig. 2). With 35 pb$^{-1}$, the result is not statistically significant enough to claim a cross-section measurement so a limit of 
$\sigma_{t\overline{t}}$ < 261 pb was set\cite{allhad}.

\begin{figure}[chi]
  \begin{minipage}[b]{0.3\linewidth}
    \centering
    \includegraphics[height=.2\textheight]{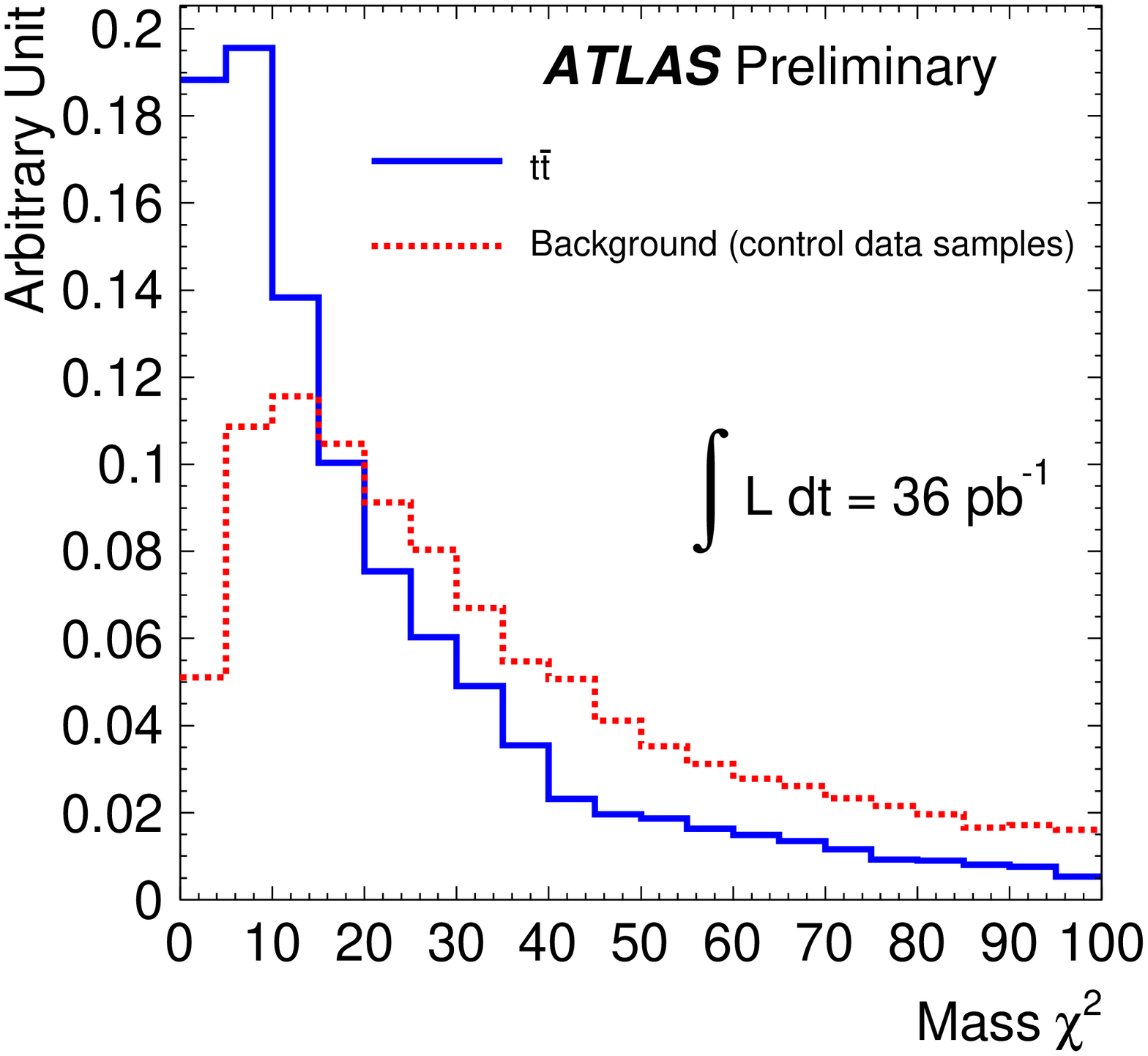}
    \caption{The caption for figure 1}
    \label{dilepfig1}
  \end{minipage}
  \hspace{0.0cm}
  \begin{minipage}[b]{0.3\linewidth}
    \centering
    \includegraphics[height=.2\textheight]{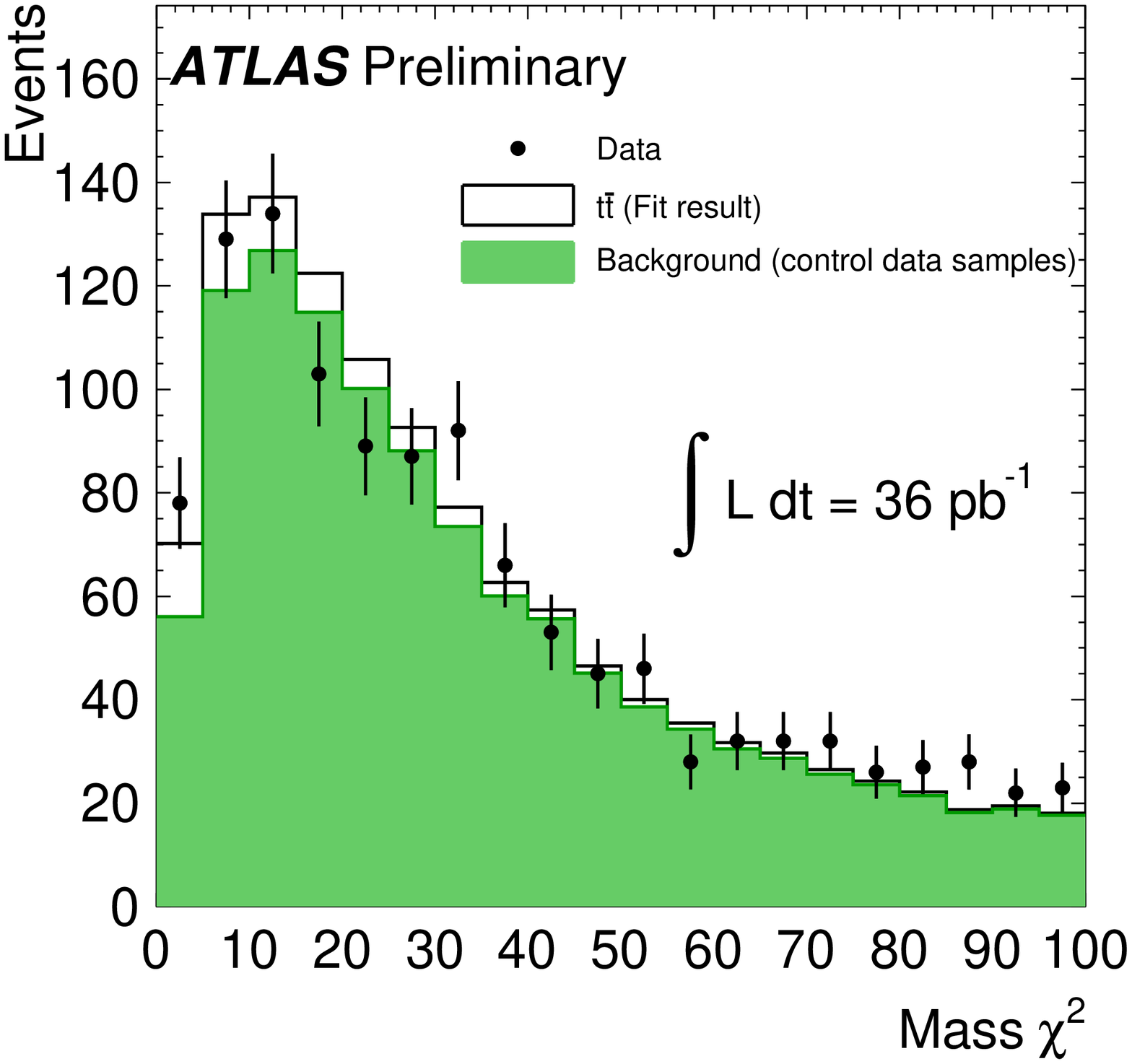}
  \caption{Distribution of the $\chi^{2}$ variable for $t\overline{t}$ and the backgrounds (left) in the all-hadronic channel.  Distribution of the $\chi^{2}$ variable for both data and Monte Carlo (right).  }
%    \caption{The number of jets in events passing the dilepton event selection (left) and the number of $b$-tagged jets (right) }
    \label{dilepfig2}
  \end{minipage}
\end{figure}

\section{Single Top}
Being an electroweak process, single top production is suppressed by the weak coupling but since only one top quark is produced, the additional phase space
means that the production rate is approximately 1/2 that of $t\overline{t}$. Experimentally, single top is very difficult to disentangle from the W+jets and 
$t\overline{t}$ backgrounds. For $t$-channel single top production, the final state contains 1 forward light quark and 1 top quark. Events that contain 2 or 3 jets, exactly 1 $b$-tagged jet, 1 lepton and ${{E}_T}^{miss}$ >25 GeV are selected as candidates for this channel. Using a cut based approach, the cross-section was
 measured to be $90^{+32}_{-22}$ pb which agrees with the Standard Model prediction of 66 pb\cite{tchan}. The $Wt$-channel has a top quark and a $W$-boson in the final
state.  The analysis was performed in the dilepton channel and vetoed events with more than 1 high $p_T$ jet. With 0.70 fb$^{-1}$ of integrated luminosity,
there is not a statistically significant signal so a limit of $\sigma_{pp \rightarrow Wt + X}$ < 39 pb was set\cite{Wtchan}.

\section{Summary and Outlook}
\begin{figure}
  \includegraphics[height=.2\textheight]{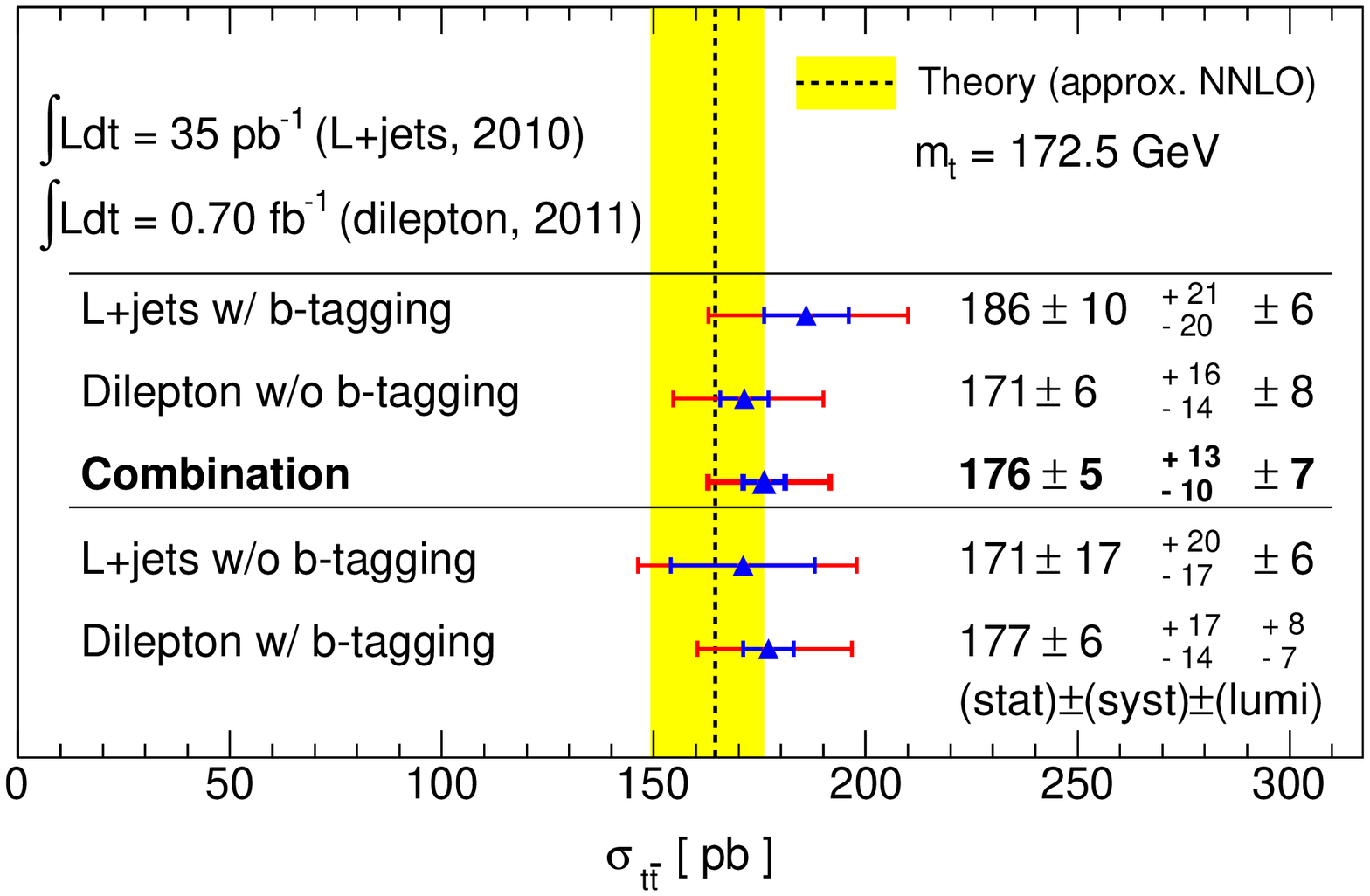}
  \caption{A summary of the lepton + jets and dilepton $t\overline{t}$ cross-section measurements\cite{combo}.}
    \label{sum}
\end{figure}

The ATLAS top quark program has made an excellent start and the outlook is bright. The excellent performance of the LHC has supplied us with enough data that many 
measurements are now limited by systematic uncertainties. By combining the lepton+jets and dilepton results, ATLAS has measured the $t\overline{t}$ cross-section to 
10\% accuracy\cite{combo} and has found it to be in agreement with the NNLO prediction as illustrated in Fig \ref{sum}.
This precision of this measurement is comparable with the 9\% uncertainty of the NNLO prediction\cite{ljetbtag}. The t-channel single top process has been measured \cite{tchan} 
and work continues in the difficult Wt and all-hadronic decay channels.  All measurements have been found to be in agreement with Standard Model
predictions. Top quark production measurements at ATLAS are entering an era of precision.

%\subsection{<A subsection>}

%\footnote{Here we test footnotes.}

%\begin{equation}
%J_{ion}=A\frac{exp\left[-\frac{E_a}{kT}\right]}{kT}\alpha \label{ionflux}
%\end{equation}

%\paragraph{<A subsubsubsection>}

\bibliographystyle{aipproc}   % if natbib is available

\bibliography{4J3_calkins}
%\begin{thebibliography}{0}

% \bibitem{atlas}
%   G.~Aad {\it et al.}  [The ATLAS Collaboration],
%   %``Expected Performance of the ATLAS Experiment - Detector, Trigger and
%   %Physics,''
%   arXiv:0901.0512 [hep-ex].
%   %%CITATION = ARXIV:0901.0512;%%
% 
% \bibitem{ljetbtag}
%     ATLAS-CONF-2011-035 \url{http://cdsweb.cern.ch/record/1337785}.
% 
% \bibitem{allhad}
%     ATLAS-CONF-2011-066 \url{http://cdsweb.cern.ch/record/1346693}.
% 
% \bibitem{dilep}
%     ATLAS-CONF-2011-100 \url{http://cdsweb.cern.ch/record/1369215}.
% 
% \bibitem{combo}
%     ATLAS-CONF-2011-108 \url{http://cdsweb.cern.ch/record/1373410}.
% 
% \bibitem{tchan}
%     ATLAS-CONF-2011-101 \url{http://cdsweb.cern.ch/record/1369217}.
% 
% \bibitem{cdf_s}
%      T.~Aaltonen {\it et al.}  [CDF Collaboration],
%   %``First Observation of Electroweak Single Top Quark Production,''
%   Phys.\ Rev.\ Lett.\  {\bf 103}, 092002 (2009)
%   [arXiv:0903.0885 [hep-ex]].
%   %%CITATION = PRLTA,103,092002;%%
% 
% \bibitem{d0_s}
%     V.~M.~Abazov {\it et al.}  [D0 Collaboration],
%   %``Observation of Single Top Quark Production,''
%   Phys.\ Rev.\ Lett.\  {\bf 103}, 092001 (2009)
%   [arXiv:0903.0850 [hep-ex]].
%   %%CITATION = PRLTA,103,092001;%%
% 
% %%not cited yet
% 
% \bibitem{ljetno}
%     ATLAS-CONF-2011-023 \url{http://cdsweb.cern.ch/record/1336753}.
% 
% \bibitem{Wtchan}
%     ATLAS-CONF-2011-104 \url{http://cdsweb.cern.ch/record/1369829}.
% 
% @Article{ATLAS-CONF-2010-999,
%  author  = "{The ATLAS Collaboration}",
%  title   = "{Place your title here}",
%  journal = "{ATLAS Note}",
%  volume  = "ATLAS-CONF-2010-999",
%  year    = "2010",
% }

%\end{thebibliography}

\IfFileExists{\jobname.bbl}{}
 {\typeout{}
  \typeout{******************************************}
  \typeout{** Please run "bibtex \jobname" to optain}
  \typeout{** the bibliography and then re-run LaTeX}
  \typeout{** twice to fix the references!}
  \typeout{******************************************}
  \typeout{}
 }

\end{document}